\documentclass[aps,twocolumn,nofootinbib,superscriptaddress,amsmath]{revtex4-1}
\usepackage{epsfig}
\usepackage{bm}
\usepackage{microtype}
\usepackage{xspace}
\usepackage{hyperref}
\usepackage{soul,color}

 \hypersetup{
   pdftitle={},%
   pdfauthor={},%
   pdfsubject={},%
   pdfkeywords={},%
   pdfstartview={},%
   bookmarksopen=true, breaklinks=true, debug=true, %
   colorlinks=true, linkcolor=red, citecolor=blue, urlcolor=blue
 }

\newcommand{\nn}{\nonumber}

\newcommand{\pythia}{{\tt Pythia8}\xspace}
\newcommand{\DDstar}{\ensuremath{\bar D^0 D^{*0}}\xspace}
\newcommand{\XYZ}{\ensuremath{XY\!Z}\xspace}
\newcommand{\X}{\ensuremath{X(3872)}\xspace}
\newcommand{\jpsi}{\ensuremath{J/\psi}\xspace}

\newcommand{\ie}{\emph{i.e.}\xspace}
\newcommand{\eg}{\emph{e.g.}\xspace}
\newcommand{\belle}{Belle}

\newcommand{\kev}{\ensuremath{{\mathrm{\,ke\kern -0.1em V}}}\xspace}
\newcommand{\mev}{\ensuremath{{\mathrm{\,Me\kern -0.1em V}}}\xspace}
\newcommand{\gev}{\ensuremath{{\mathrm{\,Ge\kern -0.1em V}}}\xspace}
\newcommand{\tev}{\ensuremath{{\mathrm{\,Te\kern -0.1em V}}}\xspace}
\newcommand{\mb}{\ensuremath{{\mathrm{\,mb}}}\xspace}
\newcommand{\fm}{\ensuremath{{\mathrm{\,fm}}}\xspace}

\begin{document}

%%%%%%%%%%%%%%%%%%%%%%%%%%%%%%%%%%%%%%%%%%%%
%% FRONTMATTER
%%%%%%%%%%%%%%%%%%%%%%%%%%%%%%%%%%%%%%%%%%%%

\title{\texorpdfstring{The nature of $X(3872)$ from high-multiplicity $pp$ collisions}{The nature of X(3872) from high-multiplicity pp collisions}}

\author{Angelo~Esposito}
\affiliation{Theoretical Particle Physics Laboratory (LPTP), Institute of Physics, EPFL, 1015 Lausanne, Switzerland}

\author{Elena~G.~Ferreiro}
\affiliation{Instituto Galego de F\'\i sica de Altas Enerx\'\i as -- IGFAE, Universidade de Santiago de Compostela, E-15782 Santiago de Compostela, Galicia-Spain}

\author{Alessandro~Pilloni}
\affiliation{European Centre for Theoretical Studies in Nuclear Physics and related Areas (ECT*) and Fondazione Bruno Kessler, Villazzano (Trento), I-38123, Italy}
\affiliation{INFN Sezione di Genova, I-16146, Genova, Italy}
\affiliation{INFN Sezione di Roma, I-00185, Roma, Italy}
\affiliation{Dipartimento di Scienze Matematiche e Informatiche, Scienze Fisiche e Scienze della Terra,
Universit\`a degli Studi di Messina, I-98166 Messina, Italy}

\author{Antonio~D.~Polosa}
\affiliation{Dipartimento di Fisica, Sapienza Universit\`a di Roma, I-00185 Roma, Italy}
\affiliation{INFN Sezione di Roma, I-00185, Roma, Italy}

\author{Carlos~A.~Salgado}
\affiliation{Instituto Galego de F\'\i sica de Altas Enerx\'\i as -- IGFAE, Universidade de Santiago de Compostela, E-15782 Santiago de Compostela, Galicia-Spain}

\begin{abstract}
The structure of exotic resonances that do not trivially fit the usual quark model expectations has been a matter of intense scientific debate during the last two decades. A possible way of estimating the size of these states is to study their behavior when immersed in QCD matter. Recently, LHCb has measured the relative abundance of the exotic \X over the ordinary $\psi(2S)$. 
We use the comover interaction model to study the yield of a compact \X. To confirm the reliability of the model in  high-multiplicity $pp$ collisions, we describe the suppression of excited over ground $\Upsilon$ states.
With this at hand, we show that the size of the compact \X  would be slightly larger than that of the $\psi(2S)$.
If the \X is instead assumed to be a meson molecule of large size, we argue that its evolution in QCD matter should be described via a coalescence model, as suggested  by data on deuteron production. We show that the predictions of this model for the \X are in contrast with data.
\end{abstract}

\maketitle

The last two decades witnessed a remarkable progress in heavy meson spectroscopy. Several new states, called \XYZ, have been observed in the quarkonium sector, close to open flavor
thresholds. Their properties are not well described by the conventional quark model/NRQCD, whence they are expected to have an exotic structure.
In particular, the \X, observed as an unexpected peak in the $\jpsi \,\pi^+ \pi^-$
invariant mass, was the first of the series~\cite{Choi:2003ue}. Its mass is almost exactly at the \DDstar threshold, and is remarkably narrow~\cite{Aaij:2020qga, Aaij:2020xjx,pdg}.\footnote{Charge conjugation is understood throughout the paper.} The pion pair is dominated by the $\rho$ meson, thus showing sizable isospin violation, unexpected if the $X$ were an ordinary charmonium. Its structure has been subject of an intense debate~\cite{Esposito:2016noz,Guo:2017jvc,Olsen:2017bmm, Brambilla:2019esw}.

Since the simple $c \bar c$ description cannot account for the observed features of \X, more valence quarks are needed. They could be aggregated by color forces in a new kind of hadron, a compact tetraquark of hadronic size $\sim 1$~fm (see e.g.~\cite{Maiani:2004vq,Brodsky:2014xia,Esposito:2016noz,Esposito:2018cwh,Braaten:2014qka}). Alternatively, {if the coupling to the closest channel is dominant,} nuclear forces could {bind} them in a hadron molecule which, given the extremely small binding energy would have a size of the order of 10\fm~\cite{Braaten:2003he, Close:2003sg, Tornqvist:2004qy, Swanson:2006st,Guo:2017jvc}, or more.

Recently, the LHCb collaboration has presented the production rates of promptly produced \X relative to the $\psi(2S)$, as a function of final state particle multiplicity~\cite{Aaij:2020hpf,*LHCb:2019obz}.
This ratio is found to decrease with increasing multiplicity,  an effect that has been known for decades to affect the production of ordinary quarkonia in proton-nucleus collisions. There is an ample consensus {for this to be}  due to final state breakup interactions of the quarkonia with comoving particles~\cite{Ferreiro:2014bia, Ferreiro:2018wbd}.

Moreover, the ALICE collaboration has recently published an analysis for deuteron production in proton-proton collisions~\cite{Acharya:2019rgc, Acharya:2020sfy}. The number of deuterons produced increases with multiplicity, hence showing a behavior that is qualitatively different from that of the \X. The idea that interactions with comovers could favor the coalescence of a hadron molecule was originally proposed in~\cite{Esposito:2013ada, Guerrieri:2014gfa,Esposito:2014rxa} for proton--proton, and in~\cite{Gyulassy:2003mc,Cho:2010db, Cho:2013rpa,Esposito:2015fsa} for nucleus--nucleus collisions. 

\begin{table*}[t]
\begin{center}\setlength{\arrayrulewidth}{0.6pt}
\begin{tabular}{c|cccc}
\hline\hline
& {$B_\mathcal{Q}$} & $r_{\cal Q}$ & $\sigma_{\cal Q}^\text{geo}$ & $\left\langle v\sigma\right\rangle_\mathcal{Q}$\\
\hline
$\psi(2S)$ & 50\mev & 0.45\fm & 6.36\mb & $4.89 \pm 0.76\mb$ \\
$X(3872)$ tetraquark  & 116\kev & 0.65\fm & 13.3\mb & $11.55 \pm 1.82\mb$\\
$X(3872)$ molecule &  116\kev & 6.6\fm & 1368\mb & $1188 \pm 187\mb$\\ 
 \hline\hline
\end{tabular}
\caption{Fixed values used in our parametrisation of the comover cross sections and the corresponding results. The procedure to compute $\left\langle v\sigma\right\rangle_{\cal Q}$ and its uncertainties are described in the text.
The {``binding energy''} is computed with respect to $D^0\bar D^0$--$D^+D^-$ for the charmonium, and with respect to \DDstar~\cite{Maiani:2017kyi} for a tetraquark \X (the  OZI-favored mode $\Lambda^+_c \Lambda_c^-$~\cite{Montanet:1980te,Rossi:2016szw,Cotugno:2009ys} being kinematically forbidden at the typical comover's energy). The average for this binding energy is $44\pm 116\kev$~\cite{Aaij:2020qga,*Aaij:2020xjx,pdg}, and in our calculations, we use the $1\sigma$ error. The radius of the tetraquark is taken from~\cite{Maiani:2017kyi,Esposito:2018cwh}. 
{The error on $\langle v\sigma\rangle_\mathcal{Q}$ depends on the uncertainty of $T_\text{eff}$, and on whether considering pionic or gluonic comovers.}}\label{tab:crosssections}
\end{center}
\end{table*}

In this work, we show how the combined study of the LHCb and ALICE points to a compact structure for the \X.

As a first step, we show that the comover model for compact states~\cite{Capella:2000zp,*Capella:2007jv,*Ferreiro:2012rq,Ferreiro:2014bia,*Ferreiro:2018wbd}  reproduces the bottomonium yields observed in high-multiplicity $pp$ collisions~\cite{Chatrchyan:2013nza}.With this at hand, we apply the same method, which assumes negligible recombination and a breakup cross section of the order of the geometrical one, to a compact tetraquark \X. The results are in agreement with the LHCb data. Pushing further the use of this model, we observe that a state of much larger size,  would get a severe suppression.

However, the case of a loosely bound, large-size, hadronic molecule is most likely described in a different way. While recombination is known to be irrelevant for compact states~\cite{Du:2015wha}, data on deuteron production in high multiplicity final states suggest the contrary. 
Therefore we extend the comover model to $i$) include the possible recombination of the hadronic pairs, and $ii)$ implement the coalescence mechanism proposed in~\cite{Esposito:2013ada,*Guerrieri:2014gfa}.
We find that, while this reproduces well the deuteron data, it fails with the \X.
Our results are therefore consistent with a compact tetraquark
interpretation, and a destruction cross section comparable to that of other compact states.

\section{The Comover Interaction Model} \label{comovers}
To include final state interactions {for compact states}, we follow the comover interaction model (CIM)~\cite{Capella:2000zp,*Capella:2007jv,*Ferreiro:2012rq,Ferreiro:2014bia,*Ferreiro:2018wbd}.
Within this framework, 
quarkonia are broken by collisions with comovers---\ie final state particles with similar rapidities. 
The density of quarkonium $\rho_{\cal Q}$, at a given transverse position~$\bm{s}$ and rapidity~$y$, for a collision of impact parameter~$\bm{b}$, evolves following
\begin{align}
\label{eq:comovrateeq}
\tau \frac{\mbox{d} \rho_{\cal Q}}{\mbox{d} \tau}\left( \bm{b},\bm{s},y \right)
= -\left\langle v \sigma\right\rangle_{\cal Q}\, \rho_{c}(\bm{b},\bm{s},y)\, \rho_{\cal Q}(\bm{b},\bm{s},y) \,,
\end{align}
where $\left\langle v \sigma\right\rangle_{\cal Q}$ is {velocity times} the cross section of quarkonium dissociation, averaged over the momentum distributions of the comoving particles, whose transverse density is $\rho_c$
at initial time $\tau_i$.
The above equation neglects recombination effects which, for a compact object, are 
irrelevant due to the paucity of heavy quarks produced in the $pp$ environment considered~\cite{Capella:2007jv,Du:2015wha}. Integrating the equation above from $\tau_i$ to $\tau_f$, we get the quarkonium density for a given position and impact parameter,
\begin{align} \label{eq:rhoQ}
\rho_{\cal Q} \propto \exp \left[-\left\langle v \sigma\right\rangle_{\cal Q} \rho_c(\bm{b},\bm{s},y)\,  \ln
\left(\frac{\rho_c(\bm{b},\bm{s},y)}{\rho_{pp} (y)}\right) \right],
\end{align}
where the argument of the logarithm comes from $\tau_f/\tau_i$ converted into ratio of densities, {effectively playing the role of time}. 
Indeed, the interaction stops at $\tau_f$, when the densities
have diluted down to $\rho_{pp}$, the value of {the minimum-bias} $pp$ density at that energy and rapidity, {as taken from~\cite{Capella:2011vi}}.

{The previous equation can already be used for an estimate of the qualitative behavior of the yields as a function of multiplicity. This can be done by neglecting the dependence on the variables and consider average values only. The comover density would simply be related to the number of charged particles by $\rho_c(\bm b,\bm s,y) \simeq \frac{3}{2}N_\text{ch}/\sigma$. The factor $3/2$ accounts for the neutral comovers, while $\sigma$ is the inelastic $pp$ cross section.}

{The full spatial dependence can be described with an eikonal-Glauber model, as it is standard in heavy ion phenomenology~\cite{Armesto:1997sa,Armesto:1998rc,Ferreiro:2012rq,Ferreiro:2014bia}. 
The basic ingredient is the profile function of the proton, taken as a Fermi function.
The comover density $\rho_c(\bm{b},\bm{s},y)$ is proportional to the number of binary parton-parton collisions per unit of transverse area {$d\bm{s}$} {and rapidity} at a given impact parameter $\bm{b}$, which in turns is proportional to the overlap between the protons---see, \eg,~\cite{Armesto:1998rc}.
The normalization is fixed to reproduce the minimum-bias $pp$ multiplicity---\ie the
$pp$ multiplicity averaged over all impact parameters.
Proceeding this way, the quarkonium yields are obtained weighting Eq.~\eqref{eq:rhoQ} with the $pp$ overlap function.}

As seen from Eq.~\eqref{eq:comovrateeq}, the quarkonium abundance is driven by its interaction cross section with comovers $\left\langle v \sigma\right\rangle_{\cal Q}$. 
In nucleus--nucleus collisions at lower energies the latter has been fitted from data (independently state by state). {However, it has also been effectively related to the geometrical cross section by}~\cite{Ferreiro:2018wbd}
{
\begin{align} \label{eq:sigma}
    \begin{split}
        \left\langle v \sigma \right\rangle_{\mathcal{Q}}&=\sigma_{\mathcal{Q}}^\text{geo} \int_{E_\mathcal{Q}^\text{thr}}^\infty dp_c \, p_c^2 \left( 1-\frac{E_\mathcal{Q}^\text{thr}}{E_c}\right)^n \mathcal{P}\big(E_c\big)\,.
    \end{split}
\end{align}
}
Here $\sigma^\text{geo}_\mathcal{Q}= \pi r_\mathcal{Q}^2$, with $r_\mathcal{Q}$ the quarkonium radius. Moreover, {$E^\text{thr}_\mathcal{Q}=B_\mathcal{Q}+m_c$, where $B_\mathcal{Q}$ is the distance between the quarkonium mass and the closest open flavor threshold (OZI-favored)}, and {$E_c=\sqrt{m_c^2+p_c^2}$} the energy of the comovers in the quarkonium center-of-mass frame. {For states close to the threshold, the breakup cross section is essentially the geometric one.}
The average is computed over {an isotropic} Bose-Einstein distribution {for the comover energy},
\begin{align}
    \mathcal{P}(E_c)\propto\frac{1}{e^{E_c/T_\text{eff}}-1}\,.
\end{align}
Both $n$ and $T_\text{eff}$ are phenomenological parameters. 
Attempts to compute $n$ using the multipole expansion in perturbative QCD at leading order suggest $n\simeq4$ for pion comovers, making the strong assumption that the scattering is initiated by the gluons inside these pions~\cite{Bhanot:1979vb,Kharzeev:1994pz,Satz:2005hx}. More realistic hadronic models suggest a smaller value~\cite{Martins:1994hd,Rapp:2008tf}. {The dependence on the comover mass (gluons or pions) is not dramatic. In Table~\ref{tab:crosssections} we quote the average of these two possibilities.} A fit on the relative yields of excited-to-ground state $\Upsilon$ data at LHC in $p\text{Pb}$ collisions gave $T_\text{eff} = 250\pm50\mev$ and $n=1$~\cite{Ferreiro:2018wbd}. We adopt the same values here.\footnote{{The values of $n$ and $T_\text{eff}$ are actually correlated~\cite{Ferreiro:2018wbd}. A difference choice for the former implies a different value of the latter, effectively compensating for the change.}}

We recall two features of the comover approach. First, larger particles are more affected by dissociation, due to  larger interaction cross sections. As a consequence, excited states are more suppressed than the ground states. 
Second, the suppression increases with comover densities, which is proportional to particle multiplicities: it increases with centrality in nucleus-nucleus collisions, and it is stronger in the nucleus direction for proton-nucleus collisions.

To confirm the applicability of the CIM also to compact states in proton--proton collisions, we use it to describe the yields of $\Upsilon$ mesons. These have been measured by CMS at 2.76\tev, as a function of the number of charged {tracks} with $p_T>400\mev${, reconstructed in the tracker at}{ and} $|\eta | < 2.4$~\cite{Chatrchyan:2013nza}.
{In Figure~\ref{fig:figppUpsilon} we show the data together with our results obtained using the breakup cross sections of~\cite{Ferreiro:2018wbd}}, confirming the validity of the model.  The global normalisation
corresponds to the experimental value at {$N_{\rm ch}=15\pm2$, that we identify with the mean multiplicity.}\footnote{{By fitting the overall normalization, we automatically account for the $p_T$ cuts of the data, which should not affect the shape dramatically~\cite{Sirunyan:2020zzb}.}}

We extend our calculation to charmonia 
by applying Eq.~\eqref{eq:sigma} for the cross sections, {using the same $T_\text{eff}$ and $n$}.
Although the non-perturbative value of $n$ could in principle be different from the bottomonium one,
we get cross section values{---$5 \mb$ for $\psi(2S)$---}in the ballpark of those obtained by directly fitting the charmonium data~\cite{Armesto:1997sa}{---$6\mb$ for $\psi(2S)$---}, confirming that a unified description is possible.

\section{The \X in the CIM model}
The relative production rates of prompt $X(3872)$ over $\psi(2S)$ have been measured by LHCb
in $pp$ collisions at 8\tev, in the forward pseudorapidity region, $2<\eta<5$~\cite{Aaij:2020hpf,*LHCb:2019obz}, as a function of the number of charged particle tracks reconstructed in the VELO detector.
This ratio is found to decrease with increasing multiplicity. 

\begin{figure}[t]
\centerline{\includegraphics[width=\columnwidth]{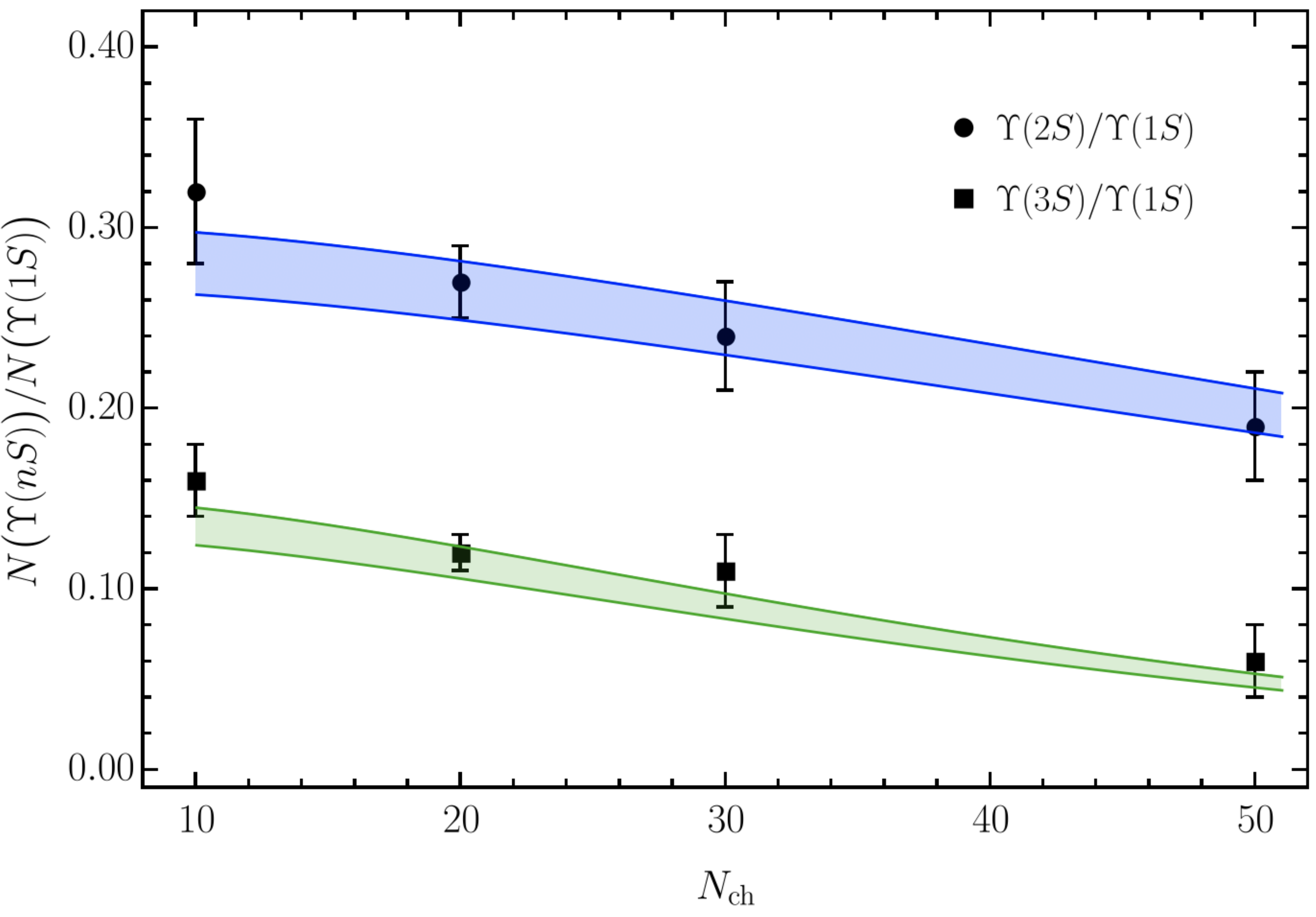}}
\caption{\label{fig:figppUpsilon}
Relative yields of
excited-to-ground state $\Upsilon$ as a function of multiplicity for $pp$ collisions at 2.76\tev in the central region, as measured by CMS~\cite{Chatrchyan:2013nza}. The bands follow the uncertainties of the {six} cross sections that contribute via the feed down, {and the one of $T_\text{eff}$}. Our results are normalised to the {experimental values corresponding to $N_\text{ch}=15\pm2$}.}
\end{figure}
\begin{figure}[t]
\includegraphics[width=\columnwidth]{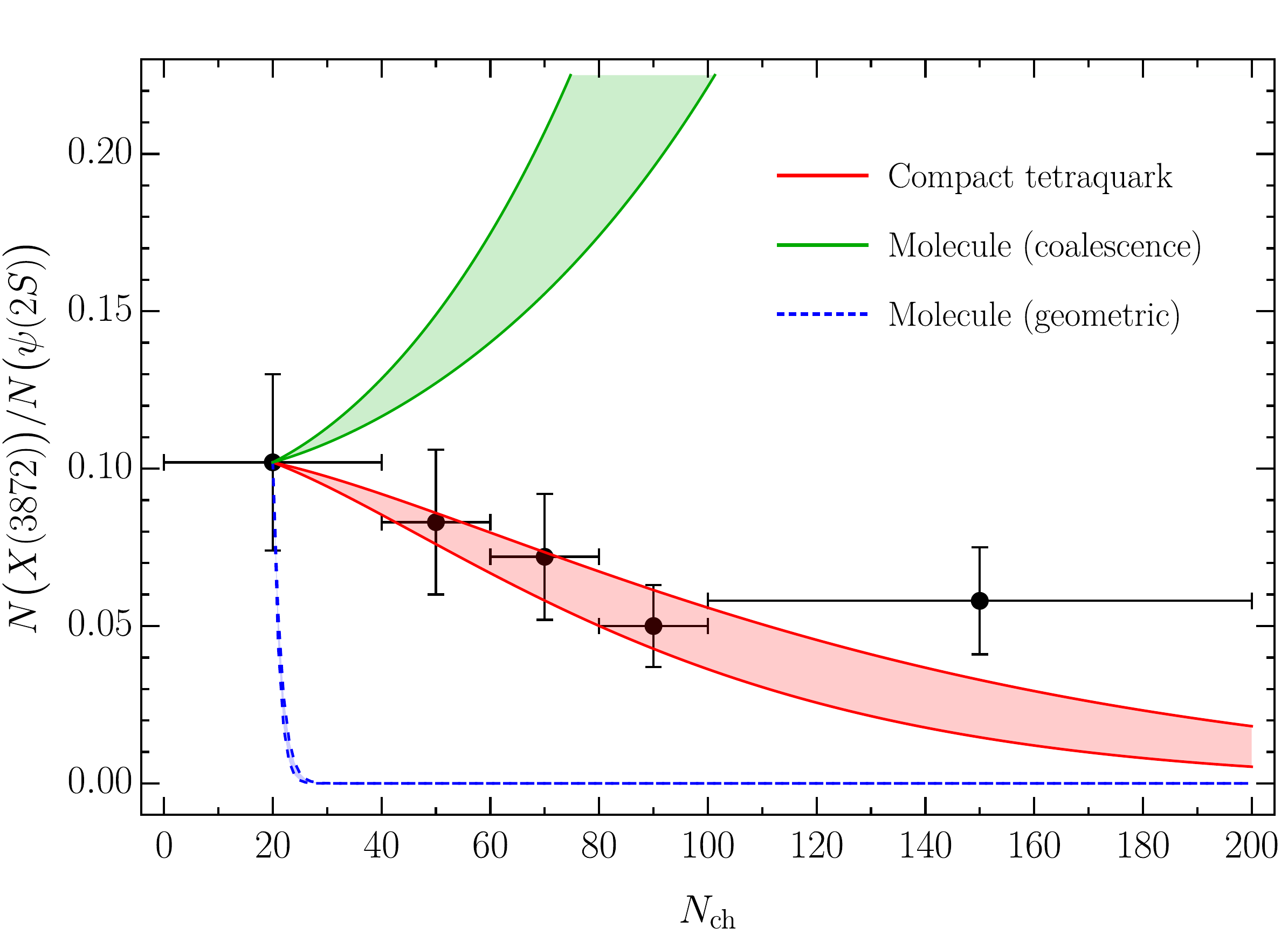}
\caption{\label{fig:figppComoversv2}
Relative yield of \X vs $\psi(2S)$ as a function of event multiplicity for $pp$ collisions at 8\tev and forward pseudorapidity, as measured by LHCb~\cite{Aaij:2020hpf,*LHCb:2019obz}. The assumption of a tetraquark of diameter 1.3\fm reproduces well the experimental data. The two error bands are computed as in Figure~\ref{fig:figppUpsilon}. Applying the geometrical picture to a molecular state predicts a very sharp suppression. The 
coalescence picture also predicts a behavior that is qualitatively different from data. See text for the definition of error band in this case.}
\end{figure}

As mentioned above, the suppression of the state is driven by its interaction cross section with the comovers, {as reported in Table~\ref{tab:crosssections}. Following the model described in the previous section, we compute the $N\big(X(3872)\big)/N\big(\psi(2S)\big)$ ratio as a function of {$N_\text{ch}$}---see Figure~\ref{fig:figppComoversv2}.} 
{For a compact tetraquark of typical hadronic size, the same physics described for quarkonia must apply, with the breakup cross section being dictated by the geometric one. Indeed, assuming a diameter of 1.3\fm~\cite{Maiani:2017kyi,Esposito:2018cwh}, the CIM gives results which describe well the LHCb data. In particular, it predicts a $~20\%$ decrease in the ratio 
when going from the first to the second multiplicity bin,
 similar to the $\Upsilon$ states (Figure~\ref{fig:figppUpsilon}).}

One could apply the same geometrical estimate for the breakup of a molecular \X. Clearly, being the size much larger, the corresponding suppression is way too strong to describe the data, as shown in Figure~\ref{fig:figppComoversv2}. The steep drop of this curve  is a qualitative indication that the description of the interaction of a molecule with comover particles must be refined. We explain this in the next section.

\section{Coalescence of hadron molecules} 
\label{sec:coalescence}

The implementation of comover interactions discussed above disregards recombination. While negligible for compact states in $pp$ collisions~\cite{Du:2015wha}, recombination is needed to explain deuteron data, whose yield  increases with increasing multiplicity of the final state~\cite{Acharya:2019rgc,*Acharya:2020sfy,Vovchenko:2018fiy,*Sun:2018mqq}. Were a molecular \X to behave like that, it would be in striking contrast with the data in~\cite{Aaij:2020hpf,*LHCb:2019obz}. {Moreover, for a  large composite object like a hadron molecule, one expects the interaction with comovers to be dominated by the scattering off the molecule constituents, an idea already put forth in a number of papers~\cite{Esposito:2013ada,*Guerrieri:2014gfa,Gyulassy:2003mc,Cho:2010db,*Cho:2013rpa,Esposito:2015fsa}.}
{A popular} description of the destruction and recombination of molecules is indeed in terms of coalescence (see, e.g.,~\cite{Kapusta:1980zz,Sato:1981ez,Gyulassy:1982pe}). In this picture the constituents are bound/free depending on whether their relative momentum is smaller/larger than some $\Lambda$. 
In~\cite{Esposito:2013ada,*Guerrieri:2014gfa} it has been proposed that the driving process 
is given by the scattering $\pi hh \rightleftharpoons \pi m$, where $hh$ are free constituents, $m$ the molecule itself and $\pi$ the comover. 

{In what follows we first present a derivation of the corresponding evolution equation. We then obtain the creation and destruction cross sections for the deuteron and the \X in the molecular regime. The effective coupling are extracted from well known hadronic physics, and are compatible with several other results (see, e.g.,~\cite{Colangelo:1994es,Matsinos:2019kqi}). The particle distributions for the molecules, free constituents and comovers are instead either taken from data, or from Monte Carlo simulations, which have been tuned over the years to reproduce hadronic results (see, e.g.,~\cite{Bignamini:2009sk}).  As a check of the soundness of our result, we note that the destruction cross section for the \X is perfectly compatible with what obtained in~\cite{Cho:2013rpa} using different techniques and a different model.
Finally, we employ all this to compute the yields of deuteron and \X as a function of the event multiplicity, which we compare with observations.}

%%%%%%%%%%%%%%%%%%%%%%%%%%%%

\subsection{Boltzmann equation for hadronic molecules and comovers} \label{app:Boltzmann}

In this derivation, we follow~\cite{Baym:1984np}. We use the standard noncovariant form of the Boltzmann equation, and the quantities discussed below are defined in the lab frame. 

Consider two hadrons, `1' and `2', with positions and momenta $(\bm{x}_i,\bm{q}_i)$. We adopt the coalescence picture, \ie that the two hadrons bind if their relative momentum is smaller than some threshold, $|\bm{q}_1-\bm{q}_2|<\Lambda$~\cite{Kapusta:1980zz,Sato:1981ez,Gyulassy:1982pe}. {In other words, if the particles are close enough in phase space, their mutual interaction will induce the formation of the molecule.} The phase-space density for these two hadrons, $n_{12}(\bm{x}_i,\bm{q}_i,\tau)$, is such that
\begin{align}
N_m(\tau)&=\int d^3x_1 d^3x_2 \int_{\mathcal{R}_\Lambda} \frac{d^3q_1}{(2\pi)^3}\frac{d^3q_2}{(2\pi)^3} \, n_{12}(\bm{x}_i,\bm{q}_i,\tau) \,, \notag \\
N_{hh}(\tau)&=\int d^3x_1 d^3x_2 \int_{\bar{\mathcal{R}}_\Lambda} \frac{d^3q_1}{(2\pi)^3}\frac{d^3q_2}{(2\pi)^3} \, n_{12}(\bm{x}_i,\bm{q}_i,\tau) \,, \notag \\
N_{12}&=\int d^3x_1 d^3x_2 \int \frac{d^3q_1}{(2\pi)^3}\frac{d^3q_2}{(2\pi)^3} \, n_{12}(\bm{x}_i,\bm{q}_i,\tau)\,,
\end{align}
where $\mathcal{R}_\Lambda$ is the domain where $|\bm{q}_1-\bm{q}_2|<\Lambda$, and $\bar{\mathcal{R}}_\Lambda$ its complement. $N_m$, $N_{hh}$ and $N_{12}$ are respectively the number of molecules, free pairs and total pairs. Clearly $N_{12} = N_m(\tau)+N_{hh}(\tau)$, and it is assumed to be constant in time, hence neglecting creation/annihilation of the constituents themselves. 
In the lab frame, the variable $\tau$ can be taken simply as the time.

We assume that the spatial and momentum distributions factorize as
\begin{align} \label{eq:split}
n_{12}\simeq\left\{\begin{array}{ll} 
\rho_m(\bm{x}_i,\tau) \, f_m(\bm{q}_i,\tau) & \!\text{ if }  |\bm{q}_1-\bm{q}_2|<\Lambda \\
\rho_{hh}(\bm{x}_i,\tau) \, f_{hh}(\bm{q}_i,\tau) & \! \text{ if } |\bm{q}_1-\bm{q}_2|\geq\Lambda
\end{array}\right.\,,
\end{align}
where both $f_m$ and $f_{hh}$ are normalized to unity when  integrated over their momentum domain. {Indeed, in absence of collisions, the particles travel freely, and the spatial and momentum distribution are hence uncorrelated. Our assumption is that, to first order in perturbation theory, this is preserved.}

Collisions with comovers change the momenta of the constituents, and consequently modify their distribution. Choosing the $z$-axis to be in the beam direction, 
{when the longitudinal expansion is much faster than the transverse one,} the Boltzmann equation reduces to~\cite{Baym:1984np}:
\begin{align} \label{eq:BB}
\begin{split}
\frac{\partial n_{12}}{\partial \tau}-\frac{q_{1z}}{\tau}\frac{\partial n_{12}}{\partial q_{1z}}-\frac{q_{2z}}{\tau}\frac{\partial n_{12}}{\partial q_{2z}}=&-L(\bm{x}_i,\bm{q}_i,\tau) \\
&+G(\bm{x}_i,\bm{q}_i,\tau)\,,
\end{split}
\end{align}
where the loss and gain terms are, respectively
\begin{subequations}
\begin{align}
L&=\int \frac{d^3q_3}{(2\pi)^3}\frac{d^3q_2'}{(2\pi)^3}\frac{d^3q_3'}{(2\pi)^3} W\big(\bm{q}_2,\bm{q}_3;\bm{q}_2',\bm{q}_3'\big)\,\delta_P \, \times \nn \\
&\qquad \times n_{12}(\bm{x}_1,\bm{x}_2,\bm{q}_1,\bm{q}_2,\tau)\,n_c(\bm{x}_2,\bm{q}_3,\tau)\,, \\
G&=\int \frac{d^3q_3}{(2\pi)^3}\frac{d^3q_2'}{(2\pi)^3}\frac{d^3q_3'}{(2\pi)^3} W\big(\bm{q}_2,\bm{q}_3;\bm{q}_2',\bm{q}_3'\big)\,\delta_P \, \times \nn \\
&\qquad \times n_{12}(\bm{x}_1,\bm{x}_2,\bm{q}_1,\bm{q}_2',\tau)\,n_c(\bm{x}_2,\bm{q}_3',\tau)\,. 
\end{align}
\end{subequations}
Here $W$ is the nonrelativistic matrix element for the $q_2+q_3\rightleftharpoons q_2'+q_3'$ process, and $\delta_P\equiv\big(2\pi\big)^4\delta^{(4)}\big(q_2+q_3-q_2'-q_3'\big)$ enforces conservation of energy and momentum. 
We assumed that the comovers interact with constituents `1' and `2' equally. Without loss of generality, we restrict the interaction to the constituent `2', so that the position of the comover must be $\bm{x}_2$. The interaction with `1' is taken into account later, by a factor of $2$ in the cross sections. Moreover, $n_c$ is the phase-space distribution of comovers, which again we factorize as $n_c(\bm{x},\bm{q},\tau)\simeq\rho_c(\bm{x},\tau)f_c(\bm{q},\tau)$. 

To study the evolution of the density of molecules, we now integrate Eq.~\eqref{eq:BB} over $\bm{q}_1$ and $\bm{q}_2$ in $\mathcal{R}_\Lambda$, {in order to isolate the molecular contribution}. 
We also assume that the momentum distribution of the molecule follows the free-stream (collisionless) distribution~\cite{Heiselberg:1998es,*Heiselberg:1995sh,*Heiselberg:1999mf}. After that, the left hand side of Eq.~\eqref{eq:BB} simply returns $\partial\rho_m(\bm{x}_i,\tau)/\partial\tau$. 
The loss term instead gives
\begin{align}
\begin{split}
\int_{\mathcal{R}_\Lambda} \frac{d^3q_1}{(2\pi)^3} \frac{d^3q_2}{(2\pi)^3} L \simeq \rho_m(\bm{x}_i,\tau)\rho_c(\bm{x}_2,\tau) \langle v\sigma\rangle_{hh}\,,
\end{split}
\end{align}
where the average cross section for the destruction of a molecule is defined as
\begin{align} \label{eq:creation}
\langle v\sigma\rangle_{hh}&\equiv\int_{\mathcal{R}_\Lambda} \frac{d^3q_1}{(2\pi)^3} \frac{d^3q_2}{(2\pi)^3}\int \frac{d^3q_3}{(2\pi)^3} \frac{d^3q_2'}{(2\pi)^3} \frac{d^3q_3'}{(2\pi
)^3} \,\times \\
&\quad\times W(\bm{q}_2,\bm{q}_3;\bm{q}_2',\bm{q}_3')\,\delta_P\, f_m(\bm{q}_1,\bm{q}_2,\tau)f_c(\bm{q}_3,\tau)\,. \notag
\end{align}
The gain term instead requires a bit more care. Imposing momentum conservation, its integral over $\mathcal{R}_\Lambda$ gives
\begin{widetext}
\begin{align}
\begin{split}
\int_{\mathcal{R}_\Lambda} \frac{d^3q_1}{(2\pi)^3} \frac{d^3q_2}{(2\pi)^3} G &= \int_{\mathcal{R}_\Lambda} \frac{d^3q_1}{(2\pi)^3} \frac{d^3q_2}{(2\pi)^3} \int \frac{d^3q_3}{(2\pi)^3}\frac{d^3q_2'}{(2\pi)^3}\frac{d^3q_3'}{(2\pi)^3} W\big(\bm{q}_2,\bm{q}_3;\bm{q}_2',\bm{q}_3'\big)\,\delta_P \times \\
&\quad \times n_{12}(\bm{x}_1,\bm{x}_2,\bm{q}_1,\bm{q}_2+\bm{q}_3-\bm{q}_3',\tau)\,n_c(\bm{x}_2,\bm{q}_3',\tau)\,. 
\end{split}
\end{align}
\end{widetext}
The relative momentum appearing in the distribution is then $\big(\bm{q}_1-\bm{q}_2\big)+\big(\bm{q}_3'-\bm{q}_3\big)$. Now, by construction $|\bm{q}_1-\bm{q}_2|<\Lambda$, while the comovers distribution in the lab frame is dominated by momenta $|\bm{q}_3'|\gg\Lambda$. This means that, barring small integration regions, for most configurations $\big|\big(\bm{q}_1-\bm{q}_2\big)+\big(\bm{q}_3'-\bm{q}_3\big)\big|\gtrsim\Lambda$, and we can use Eq.~\eqref{eq:split} and write
\begin{align}
\int_{\mathcal{R}_\Lambda} \frac{d^3q_1}{(2\pi)^3} \frac{d^3q_2}{(2\pi)^3} G \simeq \rho_{hh}(\bm{x}_i,\tau)\rho_c(\bm{x}_2,\tau) \langle v\sigma\rangle_{m}\,,
\end{align}
with average cross section for the creation of a molecule given by
\begin{align} \label{eq:destruction}
\langle v\sigma\rangle_{m}&\equiv\int_{\mathcal{R}_\Lambda} \frac{d^3q_1}{(2\pi)^3} \frac{d^3q_2}{(2\pi)^3}\int \frac{d^3q_3}{(2\pi)^3} \frac{d^3q_2'}{(2\pi)^3} \frac{d^3q_3'}{(2\pi)^3} \,\times \\
&\quad\times W(\bm{q}_2,\bm{q}_3;\bm{q}_2',\bm{q}_3')\,\delta_P\, f_{hh}(\bm{q}_1,\bm{q}_2',\tau)f_c(\bm{q}_3',\tau)\,. \notag
\end{align}
{Eqs.~\eqref{eq:creation} and \eqref{eq:destruction} are simply the cross sections for the processes of interest (destruction or creation) summed over all possible final states and averaged over the initial ones, with suitable distributions of the momenta.}
The Boltzmann equation now reads
\begin{align}
\begin{split}
\frac{\partial \rho_m(\bm{x}_i,\tau)}{\partial\tau} & \simeq \rho_{hh}(\bm{x}_i,\tau)\rho_c(\bm{x}_2,\tau)\langle v\sigma \rangle_{m} \\ 
&\quad-\rho_m(\bm{x}_i,\tau)\rho_c(\bm{x}_2,\tau)\langle v\sigma \rangle_{hh}\,.
\end{split}
\end{align}

Finally, if the density of comovers is roughly homogeneous, we can integrate over $\bm{x}_1$ and $\bm{x}_2$ as well. The final evolution equation then reads
\begin{align} \label{eq:Boltzmann}
\begin{split}
\frac{\partial N_m(\tau)}{\partial \tau} & \simeq \rho_c(\tau)\left( N_{hh}(\tau)\langle v\sigma \rangle_{m}-N_m(\tau)\langle v\sigma \rangle_{hh}\right) \\
&=\rho_c(\tau)N_{12}\langle v\sigma \rangle_{m} \\
&\quad-\rho_c(\tau)N_m(\tau)\Big(\langle v\sigma\rangle_{m}+\langle v\sigma\rangle_{hh}\Big)\,.
\end{split}
\end{align}
In the last line we used the fact that $N_{12}=N_m(\tau)+N_{hh}(\tau)$, {in order to take advantage of the fact that $N_{12}=\text{constant}$}. In the free-stream approximation, the spatial density of comovers evolves with time as $\rho_c(\tau)=\rho_c/\tau$, with $\rho_c$ the \emph{transverse} density at formation time $\tau_i$. With this approximation, 
the solution to Eq.~\eqref{eq:Boltzmann} for the number of molecules as a function of multiplicity, $N_\text{ch}$, is:
\begin{align} \label{eq:solution}
\frac{N_m}{N_{12}}&=\frac{\langle v\sigma\rangle_m}{\langle v\sigma\rangle_{hh}+\langle v\sigma\rangle_m}  +\left(\frac{N_m^0}{N_{12}} - \frac{\langle v\sigma\rangle_m}{\langle v\sigma\rangle_{hh}+\langle v\sigma\rangle_m}\right) \notag \\
&\quad \times \exp\left[{-\big(\langle v\sigma\rangle_{hh}+\langle v\sigma\rangle_m\big)\rho_c\ln(\rho_c/\rho_c^{pp})}\right]\,,
\end{align}
were $N_m^0$ is the number of molecules generated by hadronization, before any interaction with comovers. 
The dependence on multiplicity comes from the comovers spatial density. For $\langle v\sigma\rangle_m=0$ the result above reduces to Eq.~\eqref{eq:rhoQ}, upon volume integration.

%%%%%%%%%%%%%%%%%%%%%%%%%%%%%%%%%%%%%%

\subsection{Effective couplings for comover-constituent interaction} \label{app:couplings}

{To describe the creation/annihilation of the deuteron and the molecular \X by the interaction with comovers we need to determine their effective couplings.}
To simplify the computation we describe the comover-constituent scattering with a constant relativistic matrix element, $\mathcal{M}_{\pi h\to \pi h}=g^2$. As our notation suggests, we assume that the comovers are all pions, which are indeed the dominant fraction. 

For the case of the deuteron, the effective coupling is obtained by matching the total elastic pion-nucleon cross section from PDG~\cite{pdg}, averaged in the $[0,300\mev]$ kinetic energy range for the comover (therefore reaching the peak of the intermediate $\Delta$ resonance). One gets $g^2/(4\pi)\simeq 10$, which is not far from the threshold value $g^2/(4\pi)\simeq 13.5$~\cite{Matsinos:2019kqi}.

\begin{figure}
    \centering
    \includegraphics[width=.32\columnwidth]{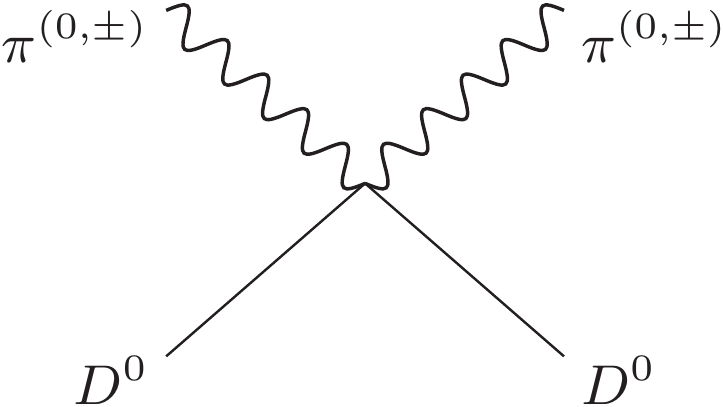}
    \includegraphics[width=.32\columnwidth]{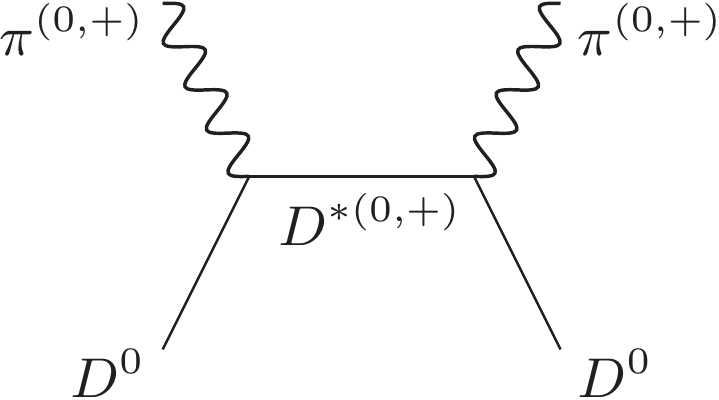}
    \includegraphics[width=.32\columnwidth]{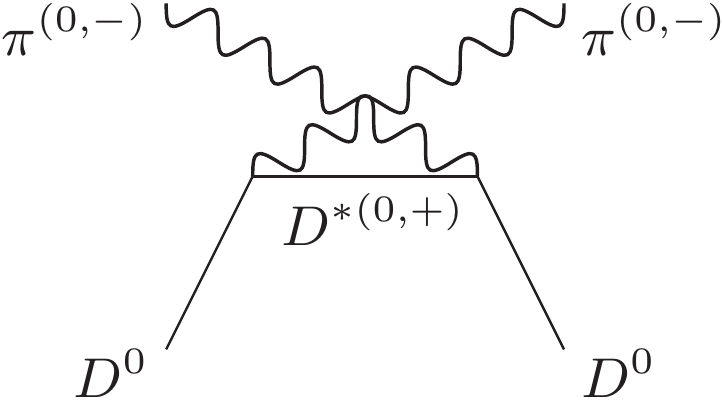}\\ \vspace{0.3cm}
    \includegraphics[width=.32\columnwidth]{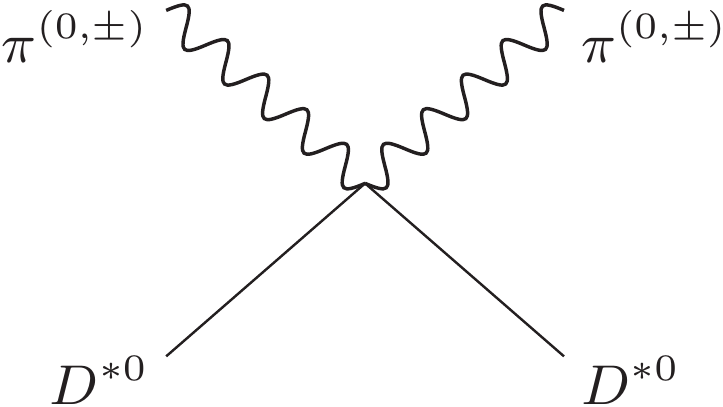}
    \includegraphics[width=.32\columnwidth]{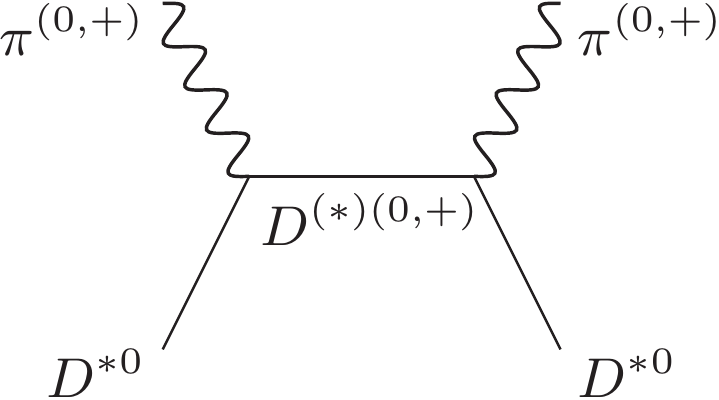}
    \includegraphics[width=.32\columnwidth]{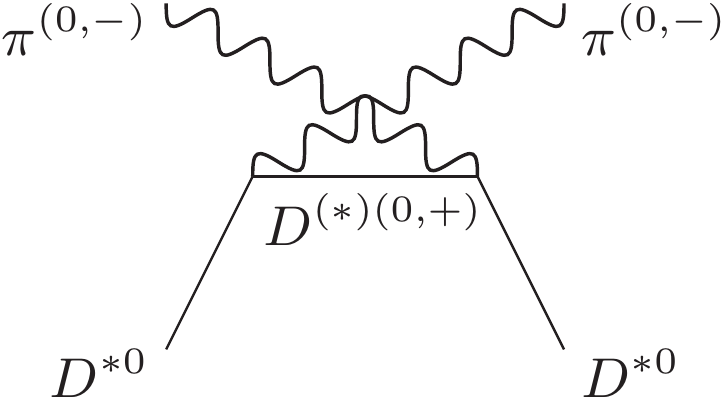}
    \caption{Feynman diagrams for the $\pi D^{(*)}\to \pi D^{(*)}$ elastic scatterings considered to extract the effective coupling $g$.}
    \label{fig:diagrams}
\end{figure}

\begin{figure}[t!]
    \centering
    \includegraphics[width=\columnwidth]{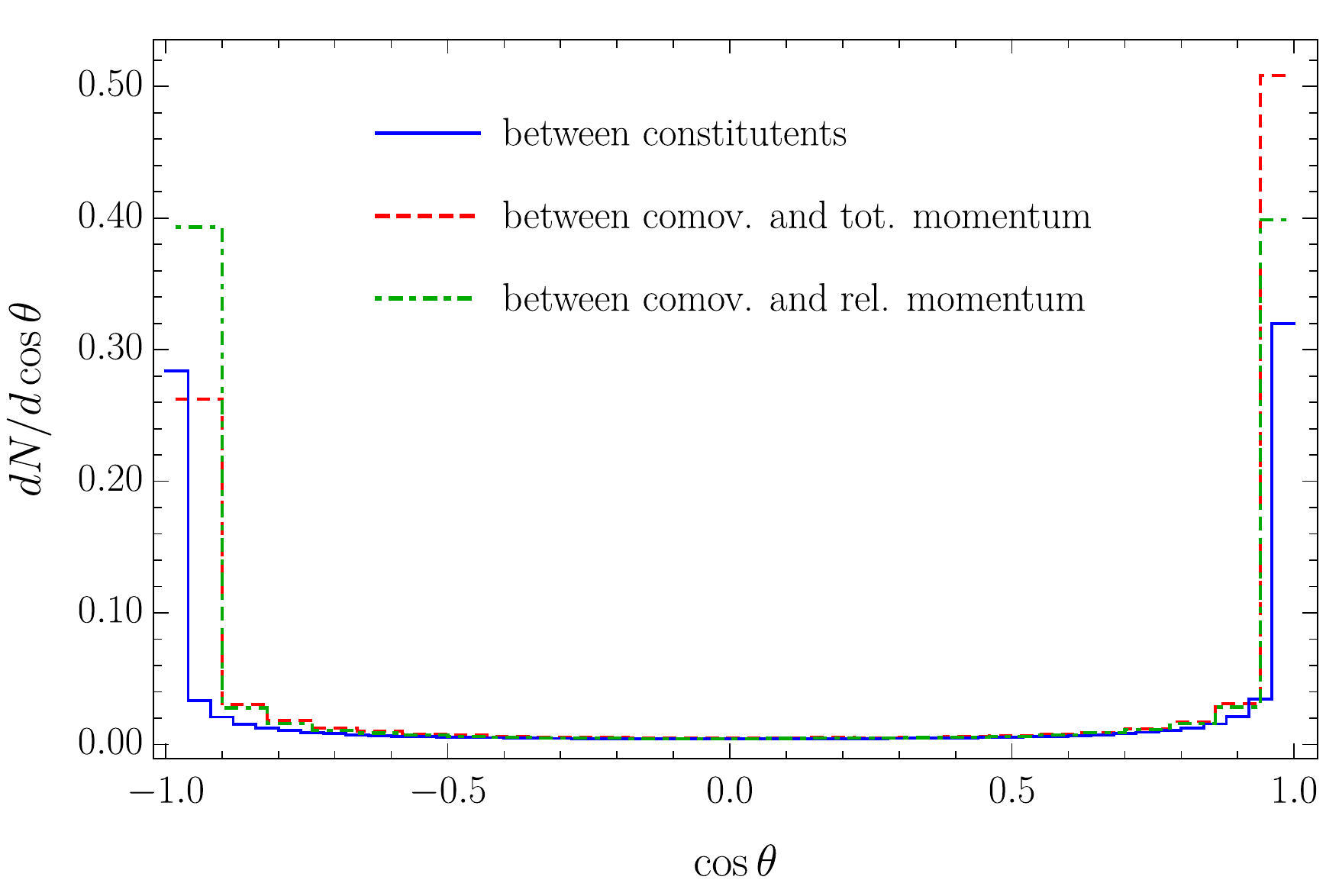}
    \caption{Distributions obtained from \pythia for the relative polar angle between the deuteron contituents (blue, solid), the comovers and the total momentum of the molecule (red, dashed) and the comovers and relative momentum of the constituents (green, dot-dashed). Everything is computed in the lab frame. Results are equivalent for the $X(3872)$.}
    \label{fig:angles}
\end{figure}

For the \X no data are available for the $\pi D^{(*)}\to\pi D^{(*)}$ scattering. To obtain the coupling we then consider the following Lagrangian for the interaction between pions and heavy mesons
\begin{align}
\begin{split}
    \mathcal{L}_\text{int}&= -\frac{y}{f_\pi}\text{tr}\left( \bar H_a H_b \gamma_\mu \gamma_5 \right) \partial^\mu \mathcal{M}_{ba} \\
    & \quad + \frac{\lambda}{4M}\text{tr}\left(\bar H_a H_a\right) \mathcal{M}_{ab}\mathcal{M}_{ba}\,.
\end{split}
\end{align}
Here $a,b$ are isospin indices, the traces are taken over Dirac matrices and $M$ is the mass of the heavy mesons. Moreover, $H_a$ is the HQET heavy meson multiplet and $\mathcal{M}_{ab}$ is the pion matrix. The trilinear coupling is given by $y\simeq0.8$~\cite{Casalbuoni:1996pg}.
We take the propagator of the $D^*$ to be
\begin{align}
\frac{-i}{p^2 - m_D^2 + i m_D \Gamma_{D^*}}\left(g^{\mu\nu} - \frac{p^\mu p^\nu}{p^2}\right)\,,
\end{align}
the real part of the pole being at the mass of the $D$ to avoid collinear divergences in the $u$-channel (note that the $D$ and $D^*$ are indeed degenerate at leading order in HQET). Moreover, we choose the projector to be exactly transverse (rather than only on-shell) to make sure that the contribution from the off-shell $D^*$ propagation vanishes at threshold. This way, the only contribution to the scattering length is given by the quartic coupling, which is found to be $\lambda\simeq 25$, by matching with lattice calculation of the $\pi D$ scattering length~\cite{Liu:2012zya}.

Given the above vertices, the effective coupling $g$ is again obtained matching the total elastic $\pi D$ cross section averaged over the kinetic energy of the comover in the $[0,300]$\mev range. The processes we considered are reported in Figure~\ref{fig:diagrams}. The result is $g^2/(4\pi)\simeq5.07$.

%%%%%%%%%%%%%%%%%%%%%%%%%%%%%%%%%%%%%%

\subsection{Creation and destruction average cross sections} \label{app:xsections}

{The last step we need to perform to implement the evolution equation~\eqref{eq:solution} is to average the cross sections over the particle distributions---see Eqs.~\eqref{eq:creation} and \eqref{eq:destruction}.}
% We describe how to extract the value of the average cross sections~\eqref{eq:creation} and \eqref{eq:destruction}. 
The calculation is conceptually straightforward but rather tedious. We will spare most of it to the reader, and only highlight the main points.

We always work in the approximation of roughly massless comovers, $m_c\simeq0$, and of loosely bound molecule, $\Lambda\ll|\bm{q}_i|$. In particular, the latter implies that the momenta of the constituents are approximately equal to each other and to half the momentum of the molecule itself.

To study the momentum distributions we employ \pythia, and generate 750k $pp$ events at $\sqrt{s}=7\tev$ for deuteron, and 2.5G $pp$ events at $8\tev$ for the \X, with full-QCD $2\to2$ matrix elements and a cut on the partonic transverse momentum, $\hat p_\perp>2\gev$. Long-lived particles are prevented from decaying. We select all the events that have at least one constituent pair in the final state and, when more than one pair per event is available, all the combinations are considered. By comovers, we mean all those particles whose momentum lies in a cone $\Delta R = \sqrt{\Delta\phi^2 + \Delta\eta^2} < 0.4$ from one of the constituents (the other constituent is clearly excluded). We checked that varying this cut within the range $\Delta R = 0.3 - 0.5$ affects the final results by no more than 3\%.

\begin{figure*}[t]
    \centering
    \includegraphics[width=\textwidth]{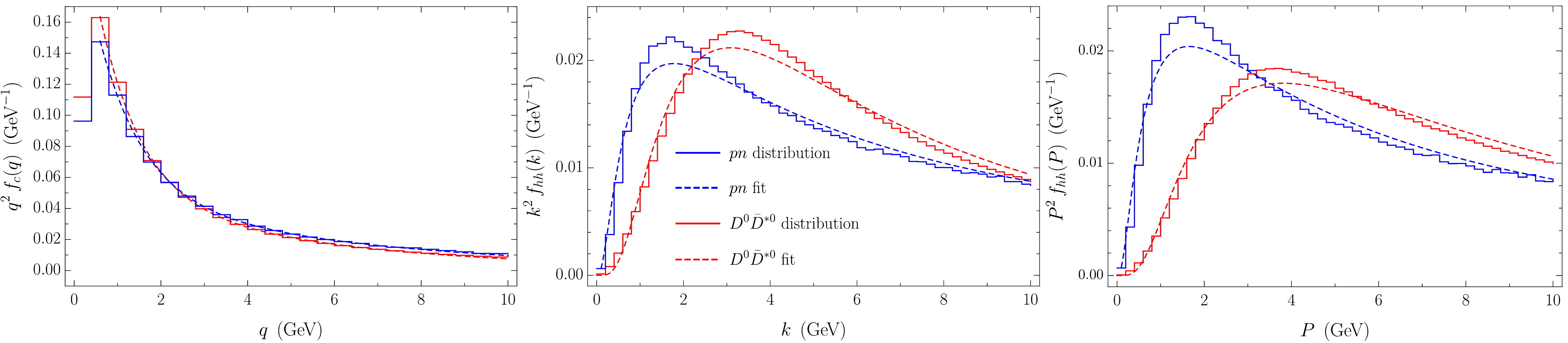}
    \caption{Distributions obtained from \pythia for the momentum of the comovers (left), the relative momentum of the hadron pair (center) and the total momentum of the pair (right) together with the best fit curves (dashed lines). Again, everything is computed in the lab frame, at $\sqrt{s}=7\tev$ for the $pn$ pairs and $\sqrt{s}=8\tev$ for the \DDstar ones.}
    \label{fig:dist}
\end{figure*}

In the lab frame, most particles have large momentum component along the beam axis. The polar angles between the constituent momenta, between the comovers and the molecule momentum, and between the comovers and the constituents relative momentum can thus be approximated by $\cos\theta\simeq\pm1$ (see Figure~\ref{fig:angles}). {This amounts to replacing the $\theta$-dependence of the corresponding distributions with a sum of $\delta$-functions with support on $\cos\theta=\pm1$.}

\begin{table*}[t]
    \centering \setlength{\arrayrulewidth}{0.6pt}
    \begin{tabular}{c|ccccccccc}
    \hline\hline
     & $\alpha_1$ & $\alpha_2$ (GeV)$^{-1}$ & $\alpha_3$ (GeV)$^{-1}$ & $\beta_1$ & $\beta_2$ (GeV)$^{-1}$ & $\beta_3$ & $\gamma_1$ & $\gamma_2$ (GeV)$^{-1}$ & $\gamma_3$  \\
     \hline\hline 
     $pn$ pairs & 3.49 & 0.17 & 0.97 & 4.12 & 3.43 & 1 & 3.99 & 2.90 & 1 \\
     \DDstar pairs & 4.04 & 0.20 & 1.03 & 7.85 & 2.09 & 2.46 & 7.33 & 1.67 & 2.84 \\
     \hline\hline
    \end{tabular}
    \caption{Best fit parameters for the distribution in Eqs.~\eqref{eq:dist} as obtained from \pythia.}
    \label{tab:parameters}
\end{table*}

After this, the two integrals can be reduced to 
\begin{widetext}
\begin{subequations} \label{eq:xsections}
\begin{align}
    \left\langle v \sigma \right\rangle_{m} &\simeq \frac{\Lambda^3 g^4}{384\pi}\int dP dk  \frac{k^2 P^2 f_{hh}\!\left(P, k \right)}{\sqrt{M^2 + \frac{\left(P + k\right)^2}{4}}\sqrt{M^2 + \frac{\left(P - k\right)^2}{4}}} \sum_{\pm} f_c\!\left(\frac{k \pm \Delta E}{2} \right) \left|\frac{k \pm \Delta E}{k \mp \Delta E}\right|\,, \\
    \left\langle v \sigma \right\rangle_{hh} &\simeq \frac{g^4}{32\pi} \int dP \frac{P^2 f_m\!\left(P\right)}{\sqrt{M^2 + \frac{P^2}{4}}}\int_{P/2}^{\infty} d q_2^\prime \frac{q_2^\prime}{\sqrt{M^2 + q_2^{\prime2}}} \int_{q_3^\text{min}}^\infty dq_3 \frac{q_3 f_c\!\left(q_3\right)}{\left(\frac{P}{2} + q_3\right)}  \theta\!\left(\sqrt{M^2+{\frac{P^2}{4}}} + q_3 \ge \sqrt{M^2 + q_2'^2}\right)\,. 
\end{align}
\end{subequations}
\end{widetext}
{Where both the sum in the first equation and the Heaviside function in the second arise from requiring for the $\delta$-functions for conservation of energy to have support on the integration region.}
Here $g$ is the comover-constituent coupling defined in section~\ref{app:couplings}, $P$ and $k$ the total and relative momenta of the hadron pair and $M$ the mass of the constituents, which we take to be approximately equal. {The momentum distributions have been integrated over all variables except those explicitly written.} Moreover, $\Delta E\equiv \sqrt{M^2+\frac{\left(P+k\right)^2}{4}} - \sqrt{M^2+\frac{\left(P-k\right)^2}{4}}$, while the minimum value of the momentum $q_3$ is
\begin{align}
    q_3^\text{min}\equiv \frac{\sqrt{4M^2+P^2}\sqrt{M^2+q_2^{\prime 2}}-2M^2-q_2^\prime P}{2q_2^\prime+\sqrt{4M^2+P^2}-P-2\sqrt{M^2+q_2^{\prime 2}}}\,.
\end{align}

The distributions extracted from \pythia are reported in Figure~\ref{fig:dist} for both the proton-neutron and the \DDstar pairs. They are well described by the following functional forms:
\begin{subequations} \label{eq:dist}
\begin{align}
    f_c(q_3) &\propto \frac{e^{-\alpha_2 q_3}+\alpha_1 e^{-\alpha_3 q_3}}{q_3^2}\,, \\
    f_{hh}(P,k) &\propto \frac{{\ln(1+\beta_2 P)}^{\beta_1}}{P^3{(1+\beta_2P)}^{\beta_3}}\frac{{\ln(1+\gamma_2 k)}^{\gamma_1}}{k^3{(1+\gamma_2k)}^{\gamma_3}}\,,
\end{align}
\end{subequations}
with best fit values given in Table~\ref{tab:parameters}. The overall constant if fixed by normalization, {as described below Eq.~\eqref{eq:split}}. The distribution of the total momentum of the molecule can instead be obtained from the experimental distributions in transverse momentum and rapidity. In particular
\begin{widetext}
\begin{align}
\begin{split}
    P^2f_m(P)&=\int dP_\perp dy \, F_m(P_\perp,y)\,\delta\!\left(P-\sqrt{P_\perp^2+(4M^2+P_\perp^2)\sinh^2y}\right)  \\
    &=\int_0^P dP_\perp \frac{2PF_m\left(P_\perp,\bar y(P_\perp)\right)}{\sqrt{4M^2+P^2}\sqrt{P^2-P_\perp^2}}\,\theta\left( \bar y(P_\perp) \leq Y \right)\,,
\end{split}
\end{align}
\end{widetext}
with $F_m(P_\perp,y)$ the experimental distribution, assumed even under $y\to-y$, $Y$ the experimental cut in rapidity, and $\sinh\bar y\equiv\sqrt{(P^2-P_\perp^2)/(4M^2+P_\perp^2)}$.

With all this at hand, we now compute the average cross sections in Eqs.~\eqref{eq:xsections}. {To test the validity of our idea, we first consider the case of the deuteron. In particular,} we take $M=938\mev$ and a coalescence momentum in the range $\Lambda=50$--$250\mev$~\cite{Ibarra:2012cc,Dal:2014nda,Aramaki:2015pii}. 
The effective coupling $g$ is discussed in section~\ref{app:couplings}. The distribution in rapidity is approximately uniform, while the one in transverse momentum is well fitted by a L\'evy-Tsallis function~\cite{Acharya:2019rgc}. We obtain $\langle v\sigma\rangle_m\simeq\left(\Lambda\big/150\mev\right)^3 \times0.51\mb$ and $\langle v\sigma\rangle_{hh}\simeq4.34\mb$.

{The suppression of the creation cross section compared to the destruction one is understood as follows. In a hadronic collision, the vast majority of constituents are produced free, with relative momentum much larger than $\Lambda$ (Figure~\ref{fig:dist}). After an interaction with a comover of momentum of order $\sim\text{GeV}$, it is unlikely for the pair to fall within the small region of phase space with $k\lesssim \Lambda$. Technically, this is due to the exponential suppression of the comover distribution at momenta of the order of the free pair ones. Similarly, if a pair is initially bound, it is much more likely for a comover to increase their relative momentum to a value larger than $\Lambda$, rather than viceversa.}

We estimate the number of initial deuterons with \pythia, by counting the proton--neutron pairs with relative momentum initially below $\Lambda$. We find $N_m^0/N_{12}=O\big(10^{-4}\big)$, which can be neglected. Hence, the dependence of the number of deuterons on multiplicity is fixed up to an overall factor, which we fit to data. We also set $\rho_{pp}$ so that the curve starts at $N_\text{ch}=1$, as in data.

\begin{figure}[t]
    \centering
    \includegraphics[width=\columnwidth]{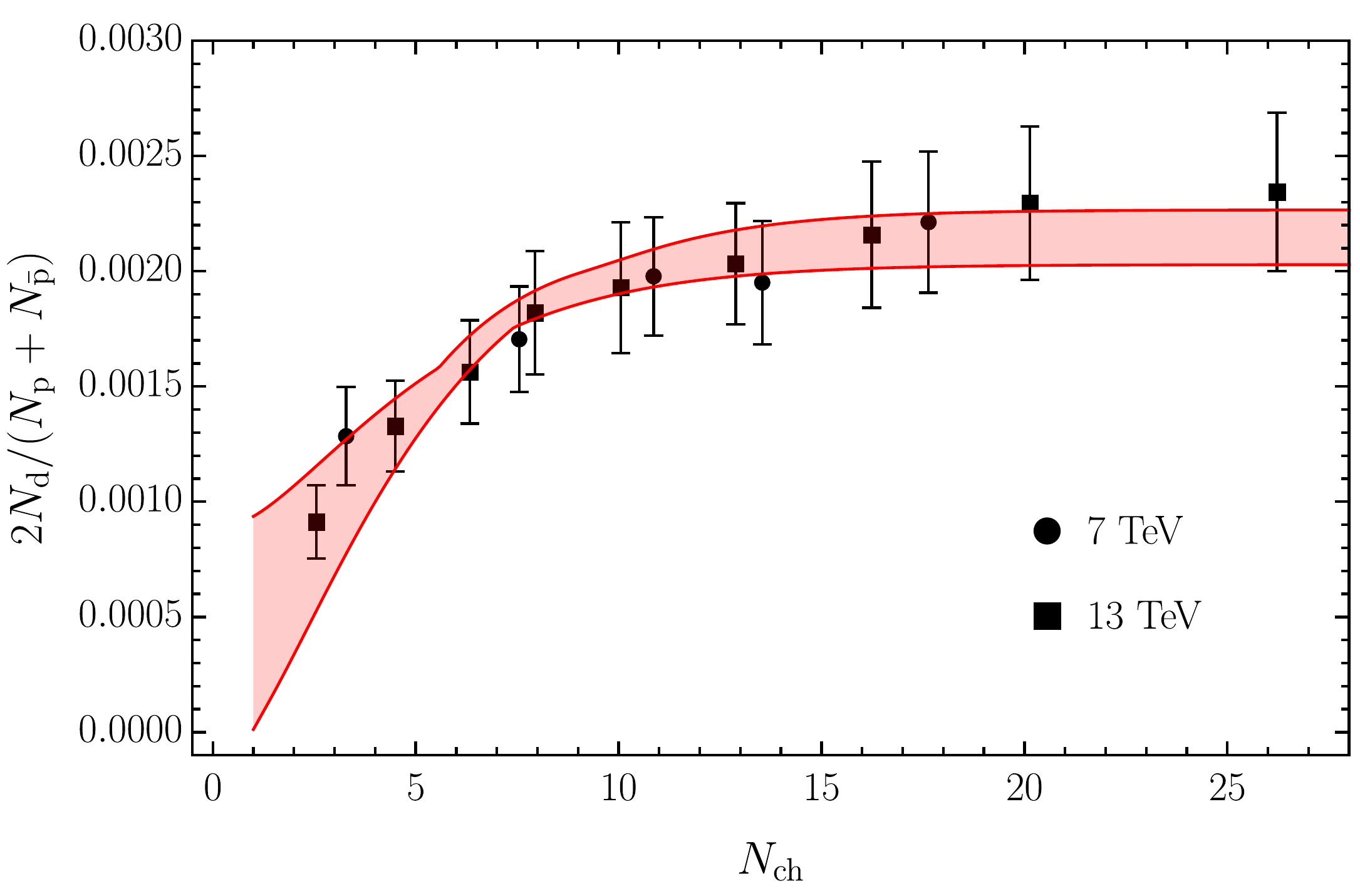}
    \caption{Number of deuterons over number of protons at 7 and 13\tev of center-of-mass energy as a function of multiplicity, as reported in~\cite{Acharya:2019rgc,Acharya:2020sfy}. The solid line is our result~\eqref{eq:solution}, the uncertainty being determined by varying the coalescence momentum $\Lambda$ between 50 and 250\mev~\cite{Ibarra:2012cc,*Dal:2014nda,*Aramaki:2015pii}.}
    \label{fig:deuteron}
\end{figure}

In Figure~\ref{fig:deuteron} we compare our results to the ALICE data. {For simplicity, we just consider the yield as in Eq.~\eqref{eq:rhoQ}, computed on the average comover density, $\rho_c(\bm{b},\bm{s},y) \simeq \frac{3}{2}N_\text{ch}/\sigma$.} The good match with experiment confirms the validity of the coalescence approach, and of the idea proposed in~\cite{Esposito:2013ada,*Guerrieri:2014gfa}, especially the relevance of the comover-constituent interaction to the enhancement of the production of hadron molecules.

\section{\texorpdfstring{The $X(3872)$ in the coalescence model}{The X(3872) in the coalescence model}}
\label{sec:Xmol}

We now apply the same procedure to the \X. The momentum distributions for the comovers and the free \DDstar pairs are again given in Eqs.~\eqref{eq:dist} and Table~\ref{tab:parameters}, while those for the \X  are obtained from a NRQCD calculation~\cite{Artoisenet:2009wk}, which reproduces well the prompt production data at high $p_\perp$~\cite{Chatrchyan:2013cld}. The coalescence momentum for the {\X} {is unknown, and we take it in the range $\Lambda=30$--$360\mev$, as proposed in the literature~\cite{Bignamini:2009sk,*Bignamini:2009fn,*Esposito:2017qef,Esposito:2013ada,Guerrieri:2014gfa,Artoisenet:2009wk}. This contributes minimally to the error band in Figure~\ref{fig:figppComoversv2}}. {With the effective coupling found in section~\ref{app:couplings}, and taking $M\simeq(m_D+m_{D^*})/2=1936\mev$, the creation and annihilation cross section are be $\langle v\sigma \rangle_m \simeq \left(\Lambda/30\mev\right)^3\times 7.1 \times 10^{-6}\mb$ and $\langle v\sigma \rangle_{hh} \simeq 0.50\mb$. The creation cross section is much smaller than the deuteron one since the constituents of the \X are produced with harder momenta (Figure~\ref{fig:dist}), and hence the creation process suffers from a stronger suppression from the exponential comover distribution: it is even more unlikely to reduce the relative momentum to a value smaller than $\Lambda$. The destruction cross section is in good agreement with what  obtained in~\cite{Cho:2013rpa} with different methods and for different processes.}

As for the initial number of \X produced by hadronization alone, $N_m^0$, there is still no consensus. On the one hand, a purely molecular interpretation requires for this number to be very small because of the difficulty in producing the constituents with such a small relative momentum~\cite{Bignamini:2009sk,*Bignamini:2009fn,*Esposito:2017qef}, as for the deuteron, albeit some controversies on final state interactions~\cite{Artoisenet:2009wk,*Artoisenet:2010uu}. On the other hand, it has been suggested that the production of the \X could be dominated by short distance physics, likely associated to a charmonium component of its wave function~\cite{Meng:2013gga,*Butenschoen:2013pxa,*Albaladejo:2017blx,*Wang:2017gay}. Were this to be true, the prompt production cross section could be significantly enhanced. Here we adopt an agnostic viewpoint and let the initial number of molecules vary from $N_m^0/N_{12}=0$ (the estimate with \pythia being $10^{-6}$) to $N_m^0/N_{12}=1$. We consider this to be the main source of uncertainty in our calculation.
Either way, the molecular nature of the \X must be manifest when propagating throughout the comovers for distances of $O(1\fm)$.

Our results are reported in Figure~\ref{fig:figppComoversv2}, and are qalitatively at odds with data, regardless on whether the molecular $X$ is copiously produced by hadronization or not. The number of \X normalized to $\psi(2S)$ always grows, {similarly to the deuteron}.
{Indeed, the typical cross sections obtained for the \X are much smaller than the one for $\psi(2S)$ shown in Table~\ref{tab:crosssections}. Thus the increasing behavior of the ratio is dominated by the decreasing $\psi(2S)$ yield, regardless of the details of $N_m^0/N_{12}$.}

\section{Conclusions}
The production of \X at high transverse momenta in low multiplicity $pp$ collisions challenges the molecular interpretation~\cite{Esposito:2015fsa, Bignamini:2009sk,*Bignamini:2009fn,*Esposito:2017qef}, to the extent that it is necessary to assume its hadronization proceeds through a  compact $c\bar c$ core.
In~\cite{Esposito:2013ada,*Guerrieri:2014gfa} it was shown that not even the interaction with comoving particles was able to account for the large number of \X, if this compact component is not considered. 
However, the recent high-multiplicity data from the LHCb and ALICE collaborations~\cite{Aaij:2020hpf,*LHCb:2019obz,Acharya:2019rgc,*Acharya:2020sfy}, encourages to reconsider the role of comovers.

In this paper we redesigned the molecule-comover interaction model, treating multiple scattering with kinetic theory. This works remarkably well at explaining the deuteron production reported by ALICE~\cite{Acharya:2019rgc,*Acharya:2020sfy}.
Were a sizeable molecular component to appear in the \X wave function, the same approach should describe its relative yield with respect to $\psi(2S)$~\cite{Aaij:2020hpf,*LHCb:2019obz}.
 However, the predicted yield always grows and cannot match the decreasing slope observed by LHCb. 
The only way to reconcile the results from the coalescence model with experiment is to make the averaged \X molecular destruction cross section, $\langle v\sigma\rangle_{hh}$, about twenty times larger {(effectively that of a compact state)}, in sharp contradiction with several agreeing determinations of the interaction couplings of pion comovers with $D,D^*$ mesons and with the findings in~\cite{Cho:2013rpa}, for example.
Our analysis concludes that, despite the closeness to the \DDstar threshold that motivated several studies (see, e.g.,~\cite{Braaten:2003he, Close:2003sg, Tornqvist:2004qy, Swanson:2006st,Guo:2017jvc}), a molecular component cannot dominate the wave function of the \X.

The LHCb results are analyzed also with the Comover Interaction Model (CIM) for compact states~\cite{Capella:2000zp,*Capella:2007jv,*Ferreiro:2012rq,Ferreiro:2014bia,*Ferreiro:2018wbd}, which is well known to describe the quarkonia yields in high multiplicity final states. The yields are determined {assuming as effective cross section the geometrical one,  \ie the size of the states}. For the first time, we apply the CIM to $pp$ collisions, and match the relative yields of $\Upsilon$ mesons with the ones reported by CMS~\cite{Chatrchyan:2013nza}. These yields decrease with multiplicity, contradicting the statement that such a behavior requires a molecular interpretation, as suggested initially for the \X. 
Conversely, it is perfectly compatible with a compact tetraquark of typical hadronic size, as we showed here.

After the appearing of this work, a modification of the CIM has been effective in reconciling the diffusion of a molecule with the LHCb data~\cite{Braaten:2020iqw}. However, we note that such description considers no recombination, which makes it hard to achieve a common understanding of the \X and deuteron data. 

We look with interest also to PbPb data~\cite{CMS:2021znk}, which seem to present novelties with respect to $pp$. It would be useful to have them binned in centrality, to allow a comparison with the deuteron data presented by ALICE~\cite{Adam:2015vda,*Acharya:2019rys}. Discussions on tetraquarks and molecules in PbPb collisions can be found in~\cite{Zhang:2020dwn,*Wu:2020zbx}.   

In conclusion, as soon as the loosely-bound molecule description is made concrete, its expected behavior appears different from what observed experimentally. To fill the discrepancy we should tune the interaction cross sections well beyond the range we estimate. In our view, the LHCb data on the \X  display the same features characterizing compact states like the $\Upsilon$ {mesons}.

%%%%%%%%%%%%%%%%%%%%%%%%%%%%%%%
%%%%%%%%%%%%%%%%%%%%%%%%%%%%%%%

\begin{acknowledgments}
E.G.F. thanks Fr\'ed\'eric Fleuret, Jean-Philippe Lansberg and Sarah Porteboeuf, who encouraged the applicability of the comover interaction to proton-proton collisions, and for fruitful discussions during the development of this work. Also, we thank Benjamin Audurier, Oscar Boente, {Eric Braaten}, Matt Durham, Feng-Kun Guo, Christoph Hanhart, Kevin Ingles and especially Luciano Maiani for valuable discussions and comments on the first version of the draft. 
We are grateful to Alexis Pompili for pointing the CMS and LHCb results presented at Quark Matter 2020 to us, and to Giuseppe Mandaglio for valuable insights on the experimental analysis by the ALICE collaboration.
A.E. is supported by the Swiss National Science Foundation under contract 200020-169696 and through the National Center of Competence in Research SwissMAP.
A.P. has received funding from the European Union's Horizon 2020 research and innovation programme under the Marie Sk{\l}odowska-Curie grant agreement No.~754496. E.G.F. and C.A.S. are supported by Ministerio de Ciencia e Innovaci\'on of Spain under project FPA2017-83814-P; Unidad de Excelencia Mar\'ia de Maetzu under project MDM-2016-0692; and Xunta de Galicia (Conseller\'ia de Educaci\'on) and FEDER. C.A.S. is supported by ERC-2018-ADG-835105 YoctoLHC.
\end{acknowledgments}

%%%%%%%%%%%%%%%%%%%%%%%%%%%%
%%%%%%%%%%%%%%%%%%%%%%%%%%%%

\bibliographystyle{apsrev4-1}
\bibliography{quattro}

 \end{document}